\documentclass[12pt,preprint]{aastex}
\usepackage{ifpdf}
\ifpdf
  \pdfoutput=1
\fi

\newcommand{\HI}{\hbox{{\sc H}{\sc i}} }
\newcommand{\NHI}{{N_{\rm HI}}}
\newcommand{\fNHI}{f(N_{\rm HI},z)}

\newcommand{\cmsq}{\,{\rm cm^{-2}}}
\newcommand{\cmcb}{\,{\rm cm^{-3}}}

\newcommand{\kb}{k_{\rm b} }
\newcommand{\fHH}{f_{\rm H_2}}   
\newcommand{\Rmol}{R_{\rm mol} }        
\newcommand{\SigHH}{\Sigma_{\rm H_2} }  
\newcommand{\SigHI}{\Sigma_{\rm HI} }   
\newcommand{\Pext}{P_{\rm ext} }        

\newcommand{\nH}{n_{\rm _{H}} }
\newcommand{\nHstar}{n^*_{\rm _{H}} }
\newcommand{\nHI}{n_{\rm _{HI}}}
\newcommand{\nHII}{n_{\rm _{HII}} }
\newcommand{\nHH}{n_{\rm _{H_2}} }
\newcommand{\xHI}{x_{\rm _{HI}} }

\newcommand{\Ob}{\Omega_{\rm b} }
\newcommand{\Om}{\Omega_{\rm m} }

\newcommand{\Ol}{\Omega_{\Lambda} }
\newcommand{\ns}{n_{\rm s} }
\newcommand{\sigeight}{\sigma_{\rm 8} }

\newcommand{\Msunh}{h^{-1} {\rm M_{\odot}} }
\newcommand{\Mpch}{h^{-1} {\rm Mpc} }
\newcommand{\kpch}{h^{-1} {\rm kpc} }
\newcommand{\pch}{h^{-1} {\rm pc} }
\newcommand{\Gadget}{{\small GADGET} }

\newcommand{\OWLS}{{\small OWLS} }

\begin{document}

\title{Through Thick and Thin - H${\hbox{\sc i}}$ Absorption in 
Cosmological Simulations}

\author
    {Gabriel Altay\altaffilmark{1}, 
      Tom Theuns\altaffilmark{1,2},  
      Joop Schaye\altaffilmark{3},
      Neil H. M. Crighton\altaffilmark{4,5} and
      Claudio Dalla Vecchia\altaffilmark{3,6}}

\email{gabriel.altay@gmail.com}
\affil{$^1$Institute for Computational Cosmology, Department of Physics,  
  Durham  University , South Road, Durham, DH1 3LE, U.K.}
\affil{$^2$Department of Physics, University of Antwerp, Campus
  Groenenborger, Groenenborgerlaan 171, B-2020 Antwerp, Belgium }
\affil{$^3$Leiden Observatory, Leiden University, P.O. Box 9513, 2300
  RA Leiden, the Netherlands}
\affil{$^4$Department of Physics, University of Durham, South Road, Durham
DH1 3LE, UK}
\affil{$^5$Max Planck Institute for Astronomy, K{\"o}nigstuhl 17, D-69117
Heidelberg, Germany}
\affil{$^6$Max-Planck-Institut f\"ur Extraterrestrische Physik, 
  Giessenbachstra\ss e, D-85478 Garching, Germany}




\begin{abstract}
We investigate the column density distribution function  of neutral 
hydrogen at redshift $z=3$ using a cosmological simulation of galaxy 
formation from the OverWhelmingly Large Simulations ({\small OWLS}) 
project.  
The base simulation includes gravity, hydrodynamics, star formation, 
supernovae feedback, stellar winds, chemodynamics, and element-by-element 
cooling in the presence of a uniform UV background. 
Self-shielding and formation of molecular hydrogen are treated
in post-processing, without introducing any free parameters, using an
accurate reverse ray-tracing algorithm and an empirical relation
between gas pressure and molecular mass fraction.  
The simulation reproduces the observed $z=3$ abundance of Ly-$\alpha$ forest, 
Lyman Limit, and Damped Ly-$\alpha$ $\HI$ absorption systems
probed by quasar sight lines over ten orders of magnitude in column density.
Self-shielding flattens the column density distribution for 
$\NHI > 10^{18} \cmsq$, while the transition to fully neutral gas and 
conversion of $\HI$ to H$_2$ steepen it around column densities of 
$\NHI = 10^{20.3} \cmsq$ and $\NHI = 10^{21.5} \cmsq$, respectively.

\end{abstract}


\section{Introduction}
\noindent
Ground-based spectroscopic observations targeting quasars are 
excellent probes of $z\geq 1.7$ neutral hydrogen 
\citep[e.g.][]{1998ARA&A..36..267R, 2005ARA&A..43..861W}.  
The Sloan Digital Sky Survey (SDSS) has produced approximately 
$1.5 \times 10^4$ moderate resolution quasar spectra 
\citep{2009ApJS..182..543A}.  These spectra provide ample 
data on $\HI$ absorption lines with column densities 
$\NHI > 10^{20.3} \cmsq$, so called 
Damped Ly-$\alpha$ systems (DLAs) 
\citep{2009ApJ...696.1543P,2009A&A...505.1087N}.  
Lines with $\NHI < 10^{17.2} \cmsq$, the so called 
Ly-$\alpha$ forest, are best discovered in high-resolution spectra of 
bright quasars \cite[e.g.][]{2002MNRAS.335..555K}. 
Lines with intermediate column densities, Lyman Limit Systems (LLSs), 
lie on the flat part of the curve of growth, which complicates the 
determination of their column densities.  Traditional methods of 
measuring $\NHI$ in DLAs can be applied to high-resolution spectra
for lines with $\NHI > 10^{19} \cmsq$ when damping wings begin to appear  
\citep[e.g.][]{2005MNRAS.363..479P, 2007ApJ...656..666O}.  
Progress  on the most difficult lines with 
$10^{14.5} \cmsq < \NHI < 10^{19} \cmsq $ has recently been made by 
\cite{2010ApJ...718..392P} by combining independent measurements of 
the Lyman limit mean free path  and integral constraints over the 
column density distribution.

Combining the observations above, one can determine the \HI column density 
distribution function $\fNHI$, {\em i.e.} the number of lines per unit 
column density $d\NHI$, per unit absorption distance $dX$, at redshifts 
$z \approx 3$ from $\NHI = 10^{12} \cmsq$ to $\NHI = 10^{22} \cmsq$.
Early determinations of $\fNHI$ at these redshifts were reasonably
well described by a single power law, 
$\fNHI \propto N_{\rm HI}^{-\eta}$, with $\eta = 1.5$
\citep{1987ApJ...321...49T}.  
As the quality of observations improved, this was no longer the case.  
\cite{1993MNRAS.262..499P}, showed that
a single power law and a double power law with a break at 
$\NHI = 10^{16} \cmsq$ both failed Kolmogorov-Smirnov tests at
the 99\% confidence level.  
The most recent
observations are well fit by a series of six power laws which 
intersect at 
$\NHI = $ 
$\{10^{14.5}$, $10^{17.3}$, $10^{19.0}$, $10^{20.3}$, $10^{21.75} \}$
$\cmsq$ \citep{2010ApJ...718..392P}.  

Attempts to explain the shape and normalization of $\fNHI$ in a
cosmological context have typically focused on sub sets of 
the full column density range.
Analytic 
\citep[e.g.][]{2001ApJ...559..507S}, 
semi-analytic \citep[e.g.][]{1997ApJ...479..523B},
and numerical 
\citep[e.g.][]{
1998MNRAS.301..478T, 
1998MNRAS.297L..49T}
models were instrumental in identifying the Ly-$\alpha$ forest 
lines with the diffuse, photo-ionized, intergalactic medium. 
Numerical work has also played a large role in determining 
properties of higher column density systems 
\citep[e.g.][]{1996ApJ...457L..57K,
1997ApJ...484...31G,
1998ApJ...495..647H,
2003ApJ...598..741C,
2004MNRAS.348..421N, 
2006ApJ...645...55R,
2007ApJ...655..685K,
2008MNRAS.390.1349P,
2009MNRAS.397..411T,
2010arXiv1008.4242H,
2010arXiv1010.5014C, 
2010ApJ...725L.219N,
2011arXiv1101.1964M}.

Although self-shielding is crucial for modelling 
optically thick absorbers, only
\cite{2006ApJ...645...55R}, 
\cite{2007ApJ...655..685K},
\cite{2008MNRAS.390.1349P}, and
\cite{2011arXiv1101.1964M} 
have used 3-D radiative transfer to calculate the attenuation of the
UV background. 
Additionally, conversion of $\HI$ to H$_2$ is thought to determine the
high end cut off in $\fNHI$
\citep{2001ApJ...562L..95S,2009ApJ...701L..12K}, yet 
only \cite{2010arXiv1010.5014C} included this process when 
modelling $\HI$ absorption. 
We present a cosmological simulation of structure
formation, to which we have applied a radiative transfer  
self-shielding calculation and a prescription for the conversion of
$\HI$ to H$_2$ without introducing any free parameters.  We show that this 
simulation reproduces observational determinations of $\fNHI$ around 
$z=3$ over the entire range in column density.  In addition, we
determine the typical neutral fractions and total hydrogen number
densities for $\HI$ absorbers as a function of column density $\NHI$. 

\begin{figure*}
\center
\epsscale{1.0}
\plotone{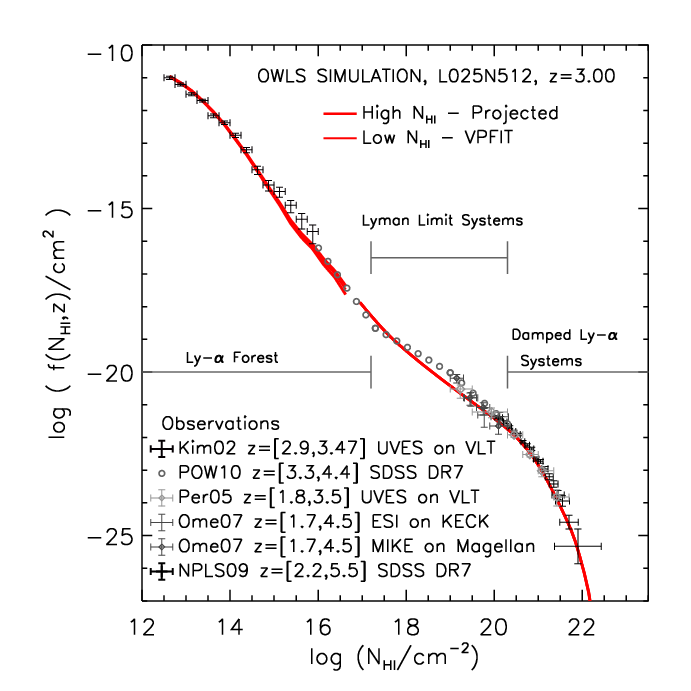}
\caption{
$\HI$ column density distribution function, $\fNHI$, at $z\sim 3$; 
simulation results are shown as curves and observational data as symbols.
The low $\NHI$ curve is obtained using mock spectra fitted with {\sc VPFIT}.
Self-shielding and H$_2$ are unimportant in this range. 
The high $\NHI$ curve is obtained by projecting the simulation
box onto  a plane and includes self-shielding and H$_2$. The gap
around $\NHI\sim 10^{17}$~cm$^{-2}$ separates low and high $\NHI$.
Poisson errors on the simulation curves are always smaller than
their thickness.  We also show high-resolution observations of the 
Ly-$\alpha$ forest  (Kim et al. 2002, ``Kim02''), 
LLSs (P{\'e}roux et al. 2005, ``Per05''; 
O'Meara et al. 2007, ``Ome07''), analysis of SDSS DLA data
(Noterdaeme et al. 2009, ``NPLS09''), and
power law constraints (Prochaska et al. 2010, ``POW10'',
open circles are spaced arbitrarily along power law segments
and do not represent $\NHI$ bins or errors).
}
\label{cddf_lrg_fig}
\end{figure*}

\section{Methodology}
We focus on model \emph{REF\_WMAP7\_L025N512} from the  
  OverWhelmingly Large Simulations ({\small OWLS}) project
  \citep{2010MNRAS.402.1536S}, which is identical to  
  \emph{REF\_L025N512} except that it was run using WMAP7
  cosmological parameters. 
This simulation was performed with a modified version of the 
Smoothed Particle Hydrodynamics (SPH) code 
\Gadget \citep{2005MNRAS.364.1105S}, and includes ``sub-grid'' models 
for star formation 
\citep{2008MNRAS.383.1210S},  
chemodynamics
\citep{2009MNRAS.399..574W}, 
galactic winds  
\citep{2008MNRAS.387.1431D}, 
and element-by-element cooling in the  presence of a uniform UV background
\citep{2009MNRAS.393...99W}.  
Gas in the interstellar medium (ISM) at densities above 
$\nHstar = 0.1\,{\rm cm^{-3}}$ is assumed to be multi-phase and star-forming.
This is modelled by imposing a polytropic equation of state (EoS) of the form  
$P = P_* (\nH/\nHstar)^{4/3}$. 
Because surface density and pressure are directly related in self-gravitating 
systems, the Kennicutt-Schmidt star formation law can be rewritten as 
a pressure law \citep{2008MNRAS.383.1210S}.  
The observed Kennicutt-Schmidt law can then be used to determine a star 
formation rate in each  gas particle.

The simulation contains $2 \times 512^3$ particles in a
periodic cube of size 25 comoving $\Mpch$.  
The cosmological parameters used are, 
 $\{\Om=0.272,\ \Ob=0.0455,\ \Ol=0.728,\ 
\sigeight=0.81,\ \ns=0.967,\ h=0.704\} $ \citep{2011ApJS..192...18K}.
The mass resolution is 
$m_{\rm b} = 1.47 \times 10^6 \Msunh$ and 
$m_{\rm dm} = 7.32 \times 10^6 \Msunh$ for baryonic and dark matter
particles, respectively.
The equivalent Plummer gravitational softening 
length is $\epsilon(z) = 1.95/(1+z)$ proper $\kpch$ at high $z$ but is not
allowed to exceed 0.5 proper $\kpch$, a value reached at $z = 2.91$.

\emph{REF\_WMAP7\_L025N512} included the \cite{2001cghr.confE..64H}
(HM01) optically thin UV background from quasars and galaxies,
but we apply a self-shielding correction  in post-processing as follows. 
For each particle, we trace rays out to a distance $l_{\rm ray}$ along 
$N_{\rm ray}$ directions defined using the HEALPix algorithm 
\citep{2005ApJ...622..759G}.  
We then compute frequency dependent optical depths along each ray and 
integrate over the HM01 spectrum to calculate a self-shielded 
photo-ionization rate, 
$\Gamma^{\rm shld}$, as opposed to the optically thin rate, 
$\Gamma^{\rm thin}_{12} = \Gamma^{\rm thin} / 10^{-12}$ s$^{-1} = 1.16$. 
This characterizes each particle with an effective optical depth 
$\tau_{\rm eff} = -\ln(\Gamma^{\rm shld}/\Gamma^{\rm thin})$.
We then use $\Gamma^{\rm shld}$ to calculate a new neutral fraction,
$\xHI=\nHI/\nH$, for each particle using an analytic equilibrium solution.
We continue to loop over the particles until the neutral fractions 
converge.  We obtain converged results for $\fNHI$ using 
$l_{\rm ray} = 100 $ proper kpc and $N_{\rm ray} = 12$.
Our self-shielding algorithm will be discussed in detail elsewhere 
(Altay et al., in prep.).

In the {\sc OWLS} snapshots, the temperature stored for gas particles on the 
polytropic star forming equation of state is simply a measure of the
imposed effective pressure.  When calculating collisional
ionization and recombination rates, we set the temperature of these
particles to $T_{\rm ISM} = 10^4$ K.  
This temperature is typical of the warm-neutral medium phase of the
ISM but our results do not change if we use lower values.  
We use case A (B) recombination rates for particles with 
$\tau_{\rm eff} < (>) 1$.
In addition, the optically thin approximation used in the
hydrodynamic simulation leads to artificial photo-heating by the UV
background in self-shielded particles.  To compensate for this, we
enforce a temperature ceiling of $T_{\rm shld} = 10^4$ K in those
particles that become self-shielded (\emph{i.e.} attain 
$\tau_{\rm eff} > 1$).  In Figure 2 we show the effect of this 
temperature correction.

For conversion of atomic hydrogen to molecules, we adopt a prescription 
based on observations by 
\cite{2006ApJ...650..933B} of 14 local spiral galaxies to form an H$_2$ 
fraction-pressure relation.  
Their sample includes various morphological types and spans a factor of five
in mean metallicity.  They obtain a power law scaling of the molecular 
fraction, $\Rmol \equiv \SigHH / \SigHI $, with the galactic mid-plane 
pressure,  $\Rmol = \left( \Pext/P_0 \right)^\alpha$, with $\alpha = 0.92$ 
and $P_0 / \kb = 3.5 \times 10^4 $ cm$^{-3}$~K. Applying this relation to the 
simulated ISM yields $\fHH = [ 1 + A
\left( \nH/\nHstar \right)^{-\beta} ]^{-1}$ with 
$A=\left( P_*/P_0 \right)^{-\alpha}$, and 
$\beta = \alpha \gamma_{\rm eff}$.

The \HI column density distribution function, 
\begin{eqnarray}
\fNHI \equiv \frac{d^2n}{d \NHI d X}
\equiv \frac{d^2n}{d \NHI d z} \frac{dz}{dX}, 
\end{eqnarray}
is defined as the number of absorption lines 
$n$, per unit column density $d \NHI$, per unit absorption distance $dX$.  
The latter is related to redshift path $dz$ as
$dX/dz = H_0 (1+z)^2/H(z)$, where $H(z)$ is the Hubble parameter  
\citep{1969ApJ...156L...7B}.
In the comparisons below, we scale $\fNHI$ reported by various 
observers to the cosmology assumed in our simulation.

The simulated $\fNHI$ below $\NHI = 10^{17} \cmsq$ is computed by
generating 1000 mock spectra through each snapshot. 
We then apply instrumental broadening with FWHM 6.6 km s$^{-1}$, add 
Gaussian noise such that we have a signal-to-noise ratio of 50 in the 
continuum, and fit the mock spectra using {\sc VPFIT} 
\citep{1987ApJ...319..709C}; see 
\cite{1998MNRAS.301..478T} for more details.
To obtain $\fNHI$ for the rarer systems with $\NHI \ge 10^{17} \cmsq$, 
we project all $512^3$ gas particles along the $z$-axis onto a grid with 
16,384$^2$ pixels using gaussian approximations to their SPH smoothing 
kernels.  
This leads to hypothetical lines of sight with a transverse spacing of 
381 proper $\pch$ or about $3/4$ the gravitational softening 
length at $z=3$.  We have verified that our results are converged with 
respect to the projected grid resolution.  
For systems with $\NHI > 10^{17.5} \cmsq $ and redshifts $z<4.4$, the
rate of incidence per unit absorption distance, $l(X)$, is observed to
be less than one \citep{2010ApJ...718..392P}.  
The absorption distance for a single sight-line through our box 
at $z=3$ is $\Delta X_1 = 0.133$ and so we expect, on average, much less than 
one system per sight-line.  Therefore, the contribution to the total
column density in projected pixels with $\NHI > 10^{17.5} \cmsq$, for the
vast majority of cases, is dominated by the single absorption system in
the line of sight.     
Curves in Figure \ref{cddf_lrg_fig} are labelled either ``VPFIT'' 
or ``Projected'', depending on the method used, all others were calculated
using projections. 

Table 1 lists the $\NHI$ bins, absorption lines per bin, and 
total absorption distance used for the low and high
$\NHI$ analyses of our fiducial model.
The $(25 \, \Mpch)^3$ volume searched for absorbers contains
$\approx 39,000$ friends of friends dark matter
halos with masses above $7.32 \times 10^8 \Msunh$ and yields 
$\approx 2 \times 10^6$ lines of sight containing DLAs.
The size of this data set obviates the need to re-weight a limited sample
of absorbers using an analytic mass function as in
\cite{1997ApJ...484...31G} or \cite{2008MNRAS.390.1349P}.

%
%
\begin{table}
  \caption{Simulation line list.}
  \label{tab:lines}
  \begin{center}
    \leavevmode
    \begin{tabular}{cc|cccc} \hline \hline              
      \multicolumn{2}{c}{{\sc VPFIT}} & \multicolumn{4}{c}{Projection} \\
      \multicolumn{2}{c}{$1000 \times \Delta X_1 = 133.1$} &
      \multicolumn{4}{c}{$16,384^2 \times \Delta X_1 = 3.574 \times 10^7$} \\
      \hline 
      $\Delta \log \NHI $ & \# of lines & 
      $\Delta \log \NHI $ & \# of lines &
      $\Delta \log \NHI $ & \# of lines \\ 
      \hline
      12.50 - 12.75  & 3598  & 17.00 - 17.10 & 858,492 & 20.00 - 20.10 & 314,774 \\
      12.75 - 13.00  & 4062  & 17.10 - 17.20 & 747,955 & 20.10 - 20.20 & 309,333 \\
      13.00 - 13.25  & 4135  & 17.20 - 17.30 & 658,685 & 20.20 - 20.30 & 302,340 \\
      13.25 - 13.50  & 3651  & 17.30 - 17.40 & 582,018 & 20.30 - 20.40 & 291,816 \\
      13.50 - 13.75  & 2918  & 17.40 - 17.50 & 518,006 & 20.40 - 20.50 & 275,818 \\
      13.75 - 14.00  & 2144  & 17.50 - 17.60 & 468,662 & 20.50 - 20.60 & 254,368 \\ 
      14.00 - 14.25  & 1362  & 17.60 - 17.70 & 431,614 & 20.60 - 20.70 & 228,520 \\
      14.25 - 14.50  & 842   & 17.70 - 17.80 & 406,575 & 20.70 - 20.80 & 198,641 \\
      14.50 - 14.75  & 466   & 17.80 - 17.90 & 387,631 & 20.80 - 20.90 & 167,671 \\
      14.75 - 15.00  & 254   & 17.90 - 18.00 & 374,532 & 20.90 - 20.00 & 135,412 \\ 
      15.00 - 15.25  & 145   & 18.00 - 18.10 & 359,789 & 21.00 - 21.10 & 103,583 \\
      15.25 - 15.50  & 73    & 18.10 - 18.20 & 350,348 & 21.10 - 21.20 & 76,751  \\
      15.50 - 15.75  & 49    & 18.20 - 18.30 & 342,146 & 21.20 - 21.30 & 54,326 \\
      15.75 - 16.00  & 40    & 18.30 - 18.40 & 334,534 & 21.30 - 21.40 & 37,745 \\ 
      16.00 - 16.25  & 25    & 18.40 - 18.50 & 329,178 & 21.40 - 21.50 & 25,140 \\
      16.25 - 16.50  & 19    & 18.50 - 18.60 & 324,411 & 21.50 - 21.60 & 16,784 \\
      16.50 - 16.75  & 11    & 18.60 - 18.70 & 320,648 & 21.60 - 21.70 & 10,938 \\
      &	                     & 18.70 - 18.80 & 318,207 & 21.70 - 21.80 & 6,740 \\
      &	                     & 18.80 - 18.90 & 316,232 & 21.80 - 21.90 & 3,667 \\
      &	                     & 18.90 - 19.00 & 314,852 & 21.90 - 22.00 & 1,614 \\
      &	                     & 19.00 - 19.10 & 314,504 & 22.00 - 22.10 & 637 \\
      &	                     & 19.10 - 19.20 & 314,583 & 22.10 - 22.20 & 206 \\
      &	                     & 19.20 - 19.30 & 313,942 & 22.20 - 22.30 & 33 \\
      &	                     & 19.30 - 19.40 & 315,802 & 22.30 - 22.40 & 14 \\
      &	                     & 19.40 - 19.50 & 316,330 & 22.40 - 22.50 & 7 \\
      &	                     & 19.50 - 19.60 & 316,884 &  &  \\
      &	                     & 19.60 - 19.70 & 317,336 &  &  \\
      &	                     & 19.70 - 19.80 & 317,979 &  &  \\
      &	                     & 19.80 - 19.90 & 316,526 &  &  \\
      &	                     & 19.90 - 20.00 & 317,212 &  &  \\
      \hline
      \hline
    \end{tabular}
  \end{center}
\end{table}

\section{Results}

\subsection{Full Range}
In Figure \ref{cddf_lrg_fig}, our fiducial model $\fNHI$ is plotted at $z=3$ 
from $\NHI=10^{12}-10^{22} \cmsq$.  
The analysis using {\sc VPFIT} in the Ly-$\alpha$ forest range, where
self-shielding and H$_2$ are not important, joins smoothly onto the
projection analysis at $\NHI > 10^{17} \cmsq$.
The model is compared to high-resolution observations of the Ly-$\alpha$ 
forest \citep{2002MNRAS.335..555K} and LLSs 
\citep{2005MNRAS.363..479P, 2007ApJ...656..666O}, DLA statistics from 
the SDSS \citep{2009A&A...505.1087N}, and a series of 
best fit power laws \citep{2010ApJ...718..392P}.

Both our model $\fNHI$ and the observations display a characteristic
flattening above the transition to Lyman Limit Systems at $\NHI =
10^{17.2} \cmsq$, and a steepening beginning around the DLA
transition, $\NHI > 10^{20.3} \cmsq$.  To quantify this model's
goodness of fit to the data, we calculate $\chi^2$ per degree of
freedom between the model and the three largest data sets using the
error bars reported by the observers.  The corresponding poisson error bars 
for our model are smaller than the thickness of the curves shown in Figure 
\ref{cddf_lrg_fig}.  
The results are 1.26 for the Ly-$\alpha$ forest data from
\cite{2002MNRAS.335..555K} and 1.26 and 2.24 for DLA data from
\cite{2009ApJ...696.1543P} and \cite{2009A&A...505.1087N},
respectively.  Lower normalizations of the UV Background as found in 
\cite{2011arXiv1105.2039H} and shown in Figure 2, would improve
these fits.

\subsection{LLSs and DLAs}

\begin{figure*}
\center
\epsscale{1.0}
\plotone{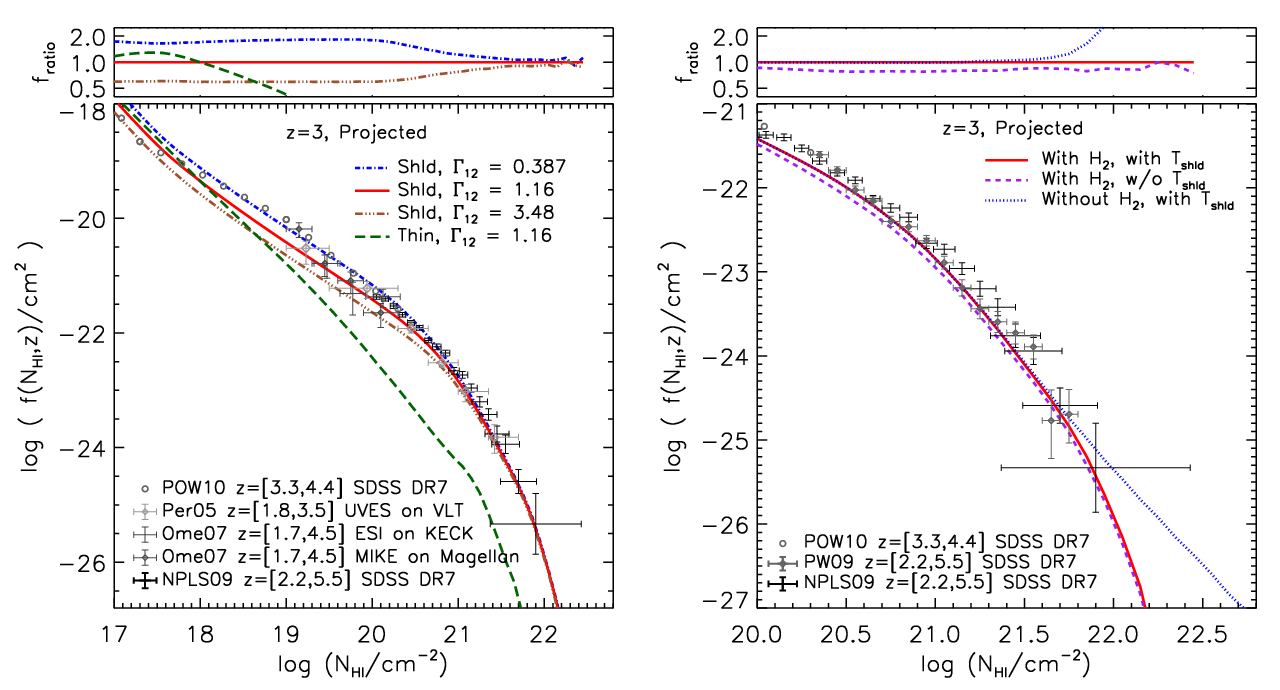}
\caption
{
$\fNHI$ - LLS and DLA range.  In the left panel, we vary the amplitude
of the UV background and show the impact of neglecting self-shielding.   
In the right panel, we isolate the effects of H$_2$ and show a model in
which we have not lowered the temperature in self-shielded particles 
(w/o $T_{\rm shld}$).    
On top of each panel, we show the ratio of each model to our fiducial model
(solid red curve), which includes self-shielding and H$_2$.  The
observational data are a subset of those in Figure \ref{cddf_lrg_fig}
plus SDSS analysis from Prochaska \& Wolfe 2009, ``PW09'', in the
right panel. 
Self-shielding becomes important for $\NHI\geq 10^{18}$~cm$^{-2}$
leading to a flattening of $\fNHI$.  
Cooling the self-shielded gas yields a constant offset while 
H$_2$ becomes important above column densities of 
$\NHI > 10^{21.5}$.} 
\label{cddf_med}
\end{figure*}

In the left panel of Figure \ref{cddf_med} we show models in which the
amplitude of the UV background was varied by factors of 3, a model with
no self-shielding, and our fiducial model.
Although systems with $\NHI > 10^{17.2} \cmsq$ are optically thick to
photons at the Lyman limit, models with and without self-shielding
(at the fiducial UV background normalization) don't diverge until 
$\NHI = 10^{18} \cmsq$.  This is because higher energy photons
with smaller cross-sections for absorption penetrate the clouds.  
Between $\NHI = 10^{17} \cmsq$ and 
$\NHI = 10^{18} \cmsq$, the model {\em with} self-shielding
predicts {\em fewer} lines,  because systems are moved to higher column
densities in the self-shielded model.  


Above $\NHI = 10^{18} \cmsq$, the model that neglects self-shielding 
stays on the Ly-$\alpha$ forest power law until it steepens around 
$\NHI = 10^{21.5} \cmsq$ due to the formation of molecules.
The other models flatten due to self-shielding and then steepen
due to both the formation of molecules and the saturation of the
neutral fraction. 
The flattening of $\fNHI$ is a hallmark of self-shielding and was also found
in the original numerical work of \cite{1996ApJ...457L..57K} and in
the analytic work of \cite{2002ApJ...568L..71Z}.
Changes in the UV background normalization by factors of three result
in constant shifts of $\fNHI$ until the gas is completely shielded
around $\NHI = 10^{21.5} \cmsq$. 
This normalization adjustment is larger than any of the uncertainties 
claimed in recent work \citep[e.g.][]{2008ApJ...688...85F}.

\subsection{DLAs}

In the right panel of Figure \ref{cddf_med}, we isolate the
effects of H$_2$, and the photo-heating of self-shielded gas. 
The models with H$_2$ approach a vertical asymptote just above 
$\NHI = 10^{22.0} \cmsq$ while the model without H$_2$ predicts the
existence of systems out to $\NHI = 10^{24.5} \cmsq$ although at such
low abundance, that less than one would have been discovered in the SDSS.


The introduction of H$_2$ produces a steepening of $\fNHI$ around
$\NHI = 10^{21.5} \cmsq$.  Such a transition, suggested theoretically in
\cite{2001ApJ...562L..95S}, has been observed at $z=0$ using CO maps as a 
tracer for H$_2$ \citep[e.g.][]{2006ApJ...643..675Z}.   
This feature coincides with the break in the double power law commonly
used to fit $\fNHI$ in the DLA column density range 
\citep{2009ApJ...696.1543P,2009A&A...505.1087N},
suggesting a relationship between the two.
At DLA column densities, ionizing radiation from local
sources may play a role \citep{2006ApJ...643...59S}.  
We have not included these sources in our self-shielding model,
but 
\cite{2010ApJ...725L.219N} have recently shown that $\fNHI$ changes by less 
than 0.1 dex when local sources are included.

Because the UV background suppresses cooling, the temperature recorded
in the \OWLS snapshots for particles that are identified as
self-shielded in post processing is an over-estimate.
To compensate for this, we enforce a temperature ceiling of 
$T_{\rm shld} = 10^4$K in self-shielded particles in our fiducial model. 
The curve labelled ``w/o $T_{\rm shld}$'' shows a model in which we have not
performed this temperature adjustment.  
Because the temperature dependence of the collisional equilibrium
neutral fraction is very small below $10^4$K, the two temperature
models should bracket the neutral fractions one would expect
from a more accurate treatment of the temperature. 
The difference between these two models is about a factor of ten 
smaller than the difference between the optically thin and
self-shielded models but can be on the order of the observational one
sigma error bars around the DLA threshold where the data are most abundant.
We plan to explore hydrodynamic simulations that include 
self-shielding in future work.

\begin{figure*}
\center
\epsscale{1.0}
\plotone{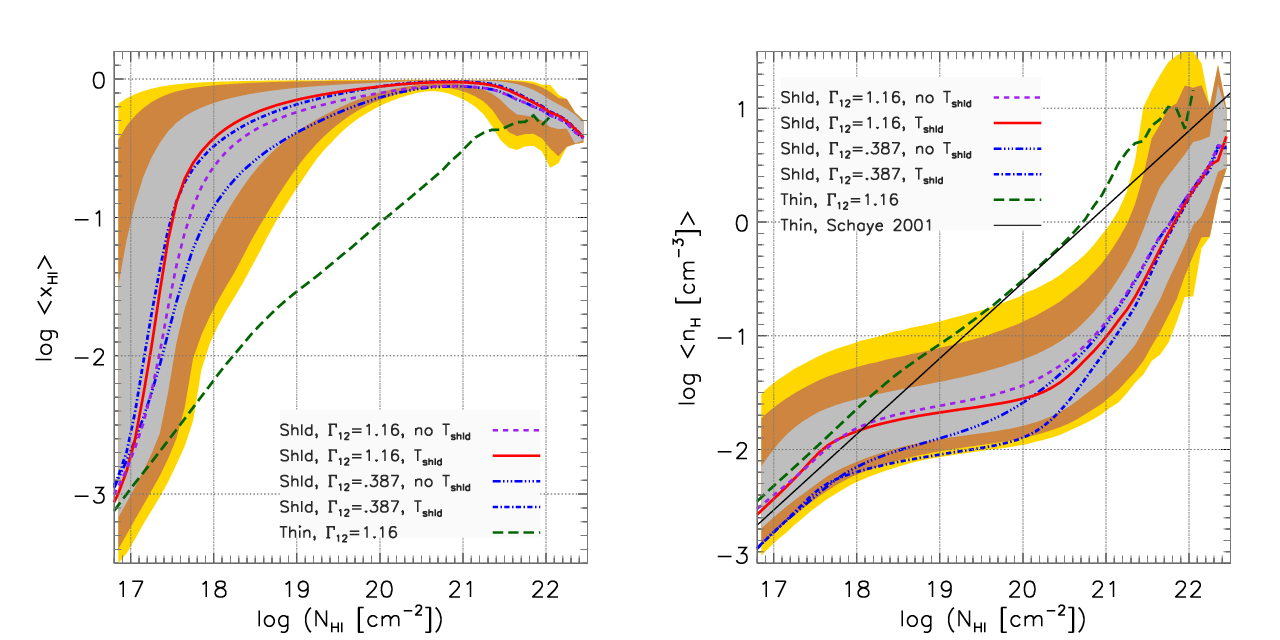}
\caption
    {\emph{Left panel}: $\nHI$ weighted neutral fraction, $\xHI(\NHI)$.  
      The red solid line indicates median values in $\NHI$ bins for the
      fiducial model which includes self-shielding and H$_2$.  
      The contours represent 68\%, 95\%, and 99\% of the data about this
      median in each bin.  
      Also shown are median values for the models shown in Figure 2
      and a model with lower UV background normalization and no
      temperature adjustment for self-shielded gas.
      \emph{Right panel}: As previous, but for the $\nHI$ weighted
      total hydrogen number density $\nH=\nHI + \nHII + 2\nHH$.  
      We also show the predictions of the analytic, optically thin
      model of \cite{2001ApJ...559..507S}.
      H$_2$ begins to reduce $\xHI$ around the DLA threshold, $\NHI =
      10^{20.3} \cmsq$ and self-shielding flattens the median $\nH$
      compared to the optically thin case between 
      $10^{18} \cmsq < \NHI < 10^{20.5} \cmsq$.  }
    \label{vsNHI}
\end{figure*}

\subsection{Physical properties of high $\NHI$ absorbers}

Neutral hydrogen mass weighted values for the neutral fraction, 
$\xHI \equiv \nHI / \nH$, and total hydrogen number density, 
$\nH = \nHI + \nHII + 2\nHH$ are plotted as a function of $\NHI$ in 
Figure \ref{vsNHI}. 
 The effects of self-shielding produce a steep deviation from the
optically thin power law in $\xHI$ above $\NHI = 10^{17} \cmsq$.
As the UV Background normalization is reduced, and as higher
temperatures are used, the deviation
becomes smaller.  The median $\xHI$ at $\NHI = 10^{18} \cmsq$ 
in our fiducial model is 0.3, however there is a large spread in the 
data in this column density range. 
It begins to drop around $\NHI = 10^{21} \cmsq$
due to the formation of H$_2$.  Systems above $\NHI = 10^{22} \cmsq$
have lost much of their atomic Hydrogen to molecules, however the H$_2$ 
likely has a small covering fraction.

The median $\nH$ flattens around the beginning of the LLS range, 
$\NHI = 10^{17.2} \cmsq$, to approximately $2 \times 10^{-2} \cmcb$
where it remains roughly constant until the start of the DLA range, 
$\NHI = 10^{20.3} \cmsq$.  
Above this column density, the gas is fully neutral (see left panel) 
causing  $\nH$ to rise steeply with $\NHI$ and $\fNHI$ to steepen 
(see Figure \ref{cddf_med}).  
Above $\NHI = 10^{21} \cmsq$, the medians for models which include
H$_2$ are steeper than linear due to the formation of molecules.  The
normalization of the UV Background and the treatment of temperature
can change the LLS characteristic density by half a decade. For
the optically thin case we find excellent agreement with the corresponding 
prediction in \cite{2001ApJ...559..507S}.

\section{Conclusions}
We have used a hydrodynamic simulation of galaxy formation together with an 
accurate ray-tracing treatment of self-shielding from the UV
background and an empirical prescription for H$_2$ formation, to compute the 
$z \approx 3$ $\HI$ column density distribution function.  
We find agreement between the reference \OWLS model and the entire
column density range probed by observations
($10^{12} \cmsq < \NHI < 10^{22} \cmsq$).  We have
shown that $\fNHI$ flattens above $\NHI = 10^{18} \cmsq$  due to
self-shielding, and steepens around $\NHI = 10^{20.3} \cmsq$ and 
$\NHI = 10^{21.5} \cmsq $ due to the absorbing gas becoming fully
neutral, and the transition from atomic to molecular hydrogen, respectively. 
In future work, we will examine the systems causing this absorption in
greater detail and repeat these analyses on a large sample of \OWLS
models.

\section*{Acknowledgments}

We thank Joseph Hennawi, Matt McQuinn, Pasquier Noterdaeme, 
Xavier Prochaska, and the \OWLS team.  These simulations were run on Stella, 
the LOFAR Blue-Gene/L system in Groningen, and on the ICC Cosmology Machine 
which is part of the DiRAC Facility jointly funded by STFC,
the Large Facilities Capital Fund of BIS, and Durham University
as part of the Virgo Consortium research programme and would not
function without the extraordinary efforts of Lydia Heck.  
This work was sponsored by the National Computing Facilities Foundation (NCF) 
for the use of supercomputer facilities, with financial support from the 
Netherlands Organization for Scientific Research (NWO).

\label{lastpage}


\begin{thebibliography}{50}
\expandafter\ifx\csname natexlab\endcsname\relax\def\natexlab#1{#1}\fi


\bibitem[Abazajian et al.(2009)]{2009ApJS..182..543A} Abazajian, K.~N., et 
al.\ 2009, \apjs, 182, 543 

\bibitem[Bahcall \& Peebles(1969)]{1969ApJ...156L...7B} 
Bahcall, J.~N., \& Peebles, P.~J.~E.\ 1969, \apjl, 156, L7 

\bibitem[{Bi} \& {Davidsen}(1997)]{1997ApJ...479..523B}
{Bi} H., {Davidsen} A.~F., 1997, \apj, 479, 523

\bibitem[{Blitz} \& {Rosolowsky}(2006)]{2006ApJ...650..933B}
{Blitz} L., {Rosolowsky} E., 2006, \apj, 650, 933

\bibitem[{Carswell} et~al.(1987){Carswell}, {Webb}, {Baldwin} \&
  {Atwood}]{1987ApJ...319..709C}
{Carswell} R.~F., {Webb} J.~K., {Baldwin} J.~A., {Atwood} B., 1987, \apj, 319,
  709


\bibitem[{Cen} et~al.(2003){Cen}, {Ostriker}, {Prochaska} \&
  {Wolfe}]{2003ApJ...598..741C}
{Cen} R., {Ostriker} J.~P., {Prochaska} J.~X., {Wolfe} A.~M., 2003, \apj, 598,
  741

\bibitem[Cen(2010)]{2010arXiv1010.5014C} Cen, R.\ 2010, arXiv:1010.5014 

\bibitem[{Dalla Vecchia} \& {Schaye}(2008)]{2008MNRAS.387.1431D}
{Dalla Vecchia} C., {Schaye} J., 2008, \mnras, 387, 1431


\bibitem[{Faucher-Gigu{\`e}re} et~al.(2008){Faucher-Gigu{\`e}re}, {Lidz},
  {Hernquist} \& {Zaldarriaga}]{2008ApJ...688...85F}
{Faucher-Gigu{\`e}re} C., {Lidz} A., {Hernquist} L., {Zaldarriaga} M., 2008,
  \apj, 688, 85


\bibitem[{Gardner} et~al.(1997){Gardner}, {Katz}, {Hernquist} \&
  {Weinberg}]{1997ApJ...484...31G}
{Gardner} J.~P., {Katz} N., {Hernquist} L., {Weinberg} D.~H.,
  1997, \apj, 484, 31

\bibitem[{G{\'o}rski} et~al.(2005){G{\'o}rski}, {Hivon}, {Banday}
  et~al.]{2005ApJ...622..759G}
{G{\'o}rski} K.~M., {Hivon} E., {Banday} A.~J., et~al., 2005, \apj, 622, 759

\bibitem[{Haardt} \& {Madau}(2001)]{2001cghr.confE..64H}
{Haardt} F., {Madau} P., 2001, in { Clusters of Galaxies and the High Redshift
  Universe Observed in X-rays\/}, edited by {D.~M.~Neumann \& J.~T.~V.~Tran}

\bibitem[Haardt \& Madau(2011)]{2011arXiv1105.2039H} 
Haardt, F., \& Madau, P.\ 2011, arXiv:1105.2039 




\bibitem[Haehnelt et al.(1998)]{1998ApJ...495..647H} Haehnelt, M.~G., 
Steinmetz, M., \& Rauch, M.\ 1998, \apj, 495, 647 


\bibitem[Hong et al.(2010)]{2010arXiv1008.4242H} Hong, S., Katz, N., 
Dav{\'e}, R., Fardal, M., Kere{\v s}, D., 
\& Oppenheimer, B.~D.\ 2010, arXiv:1008.4242 


\bibitem[{Katz} et~al.(1996){Katz}, {Weinberg}, {Hernquist} \&
  {Miralda-Escude}]{1996ApJ...457L..57K}
{Katz} N., {Weinberg} D.~H., {Hernquist} L., {Miralda-Escude} J., 1996, \apjl,
  457, L57+

\bibitem[{Kim} et~al.(2002){Kim}, {Carswell}, {Cristiani}, {D'Odorico} \&
  {Giallongo}]{2002MNRAS.335..555K}
{Kim} T., {Carswell} R.~F., {Cristiani} S., {D'Odorico} S., {Giallongo} E.,
  2002, \mnras, 335, 555

\bibitem[Kohler \& Gnedin(2007)]{2007ApJ...655..685K} 
Kohler, K., \& Gnedin, N.~Y.\ 2007, \apj, 655, 685 


\bibitem[Komatsu et al.(2011)]{2011ApJS..192...18K} Komatsu, E., et al.\ 
2011, \apjs, 192, 18 

\bibitem[{Krumholz} et~al.(2009){Krumholz}, {Ellison}, {Prochaska} \&
  {Tumlinson}]{2009ApJ...701L..12K}
{Krumholz} M.~R., {Ellison} S.~L., {Prochaska} J.~X., {Tumlinson} J., 2009,
  \apjl, 701, L12

\bibitem[{McQuinn} et~al.(2011)]{2011arXiv1101.1964M} 
{McQuinn} M., {Oh} S.~P., {Faucher-Giguere} C.-A., 2011,
arXiv:1101.1964


\bibitem[Nagamine et al.(2010)]{2010ApJ...725L.219N} Nagamine, K., Choi, 
J.-H., \& Yajima, H.\ 2010, \apjl, 725, L219 


\bibitem[{Nagamine} et~al.(2004){Nagamine}, {Springel} \&
 {Hernquist}]{2004MNRAS.348..421N}
{Nagamine} K., {Springel} V., {Hernquist} L., 2004, \mnras, 348, 421

\bibitem[{Noterdaeme} et~al.(2009){Noterdaeme}, {Petitjean}, {Ledoux} \&
  {Srianand}]{2009A&A...505.1087N}
{Noterdaeme} P., {Petitjean} P., {Ledoux} C., {Srianand} R., 2009, \aap, 505,
  1087

\bibitem[{O'Meara} et~al.(2007){O'Meara}, {Prochaska}, {Burles}, {Prochter},
  {Bernstein} \& {Burgess}]{2007ApJ...656..666O}
{O'Meara} J.~M., {Prochaska} J.~X., {Burles} S., {Prochter} G., {Bernstein}
  R.~A., {Burgess} K.~M., 2007, \apj, 656, 666

\bibitem[{P{\'e}roux} et~al.(2005){P{\'e}roux}, {Dessauges-Zavadsky},
  {D'Odorico}, {Sun Kim} \& {McMahon}]{2005MNRAS.363..479P}
{P{\'e}roux} C., {Dessauges-Zavadsky} M., {D'Odorico} S., {Sun Kim} T.,
  {McMahon} R.~G., 2005, \mnras, 363, 479

\bibitem[{Petitjean} et~al.(1993){Petitjean}, {Webb}, {Rauch}, {Carswell} \&
  {Lanzetta}]{1993MNRAS.262..499P}
{Petitjean} P., {Webb} J.~K., {Rauch} M., {Carswell} R.~F., {Lanzetta} K.,
  1993, \mnras, 262, 499

\bibitem[{Pontzen} et~al.(2008){Pontzen}, {Governato}, {Pettini}
  et~al.]{2008MNRAS.390.1349P}
{Pontzen} A., {Governato} F., {Pettini} M., et~al., 2008, \mnras, 390, 1349

\bibitem[{Prochaska} et~al.(2010){Prochaska}, {O'Meara} \&
  {Worseck}]{2010ApJ...718..392P}
{Prochaska} J.~X., {O'Meara} J.~M., {Worseck} G., 2010, \apj, 718, 392

\bibitem[{Prochaska} \& {Wolfe}(2009)]{2009ApJ...696.1543P}
{Prochaska} J.~X., {Wolfe} A.~M., 2009, \apj, 696, 1543

\bibitem[{Rauch}(1998)]{1998ARA&A..36..267R}
{Rauch} M., 1998, \araa, 36, 267

\bibitem[Razoumov et al.(2006)]{2006ApJ...645...55R} Razoumov, A.~O., 
Norman, M.~L., Prochaska, J.~X., \& Wolfe, A.~M.\ 2006, \apj, 645, 55 

\bibitem[{Schaye}(2001{\natexlab{a}})]{2001ApJ...559..507S}
{Schaye} J., 2001{\natexlab{a}}, \apj, 559, 507

\bibitem[{Schaye}(2001{\natexlab{b}})]{2001ApJ...562L..95S}
{Schaye} J., 2001{\natexlab{b}}, \apjl, 562, L95

\bibitem[Schaye(2006)]{2006ApJ...643...59S} 
Schaye, J.\ 2006, \apj, 643, 59 

\bibitem[{Schaye} \& {Dalla Vecchia}(2008)]{2008MNRAS.383.1210S}
{Schaye} J., {Dalla Vecchia} C., 2008, \mnras, 383, 1210

\bibitem[{Schaye} et~al.(2010){Schaye}, {Dalla Vecchia}, {Booth}
  et~al.]{2010MNRAS.402.1536S}
{Schaye} J., {Dalla Vecchia} C., {Booth} C.~M., et~al., 2010, \mnras, 402, 1536

\bibitem[{Springel}(2005)]{2005MNRAS.364.1105S}
{Springel} V., 2005, \mnras, 364, 1105

\bibitem[{Tescari} et~al.(2009){Tescari}, {Viel}, {Tornatore} \&
  {Borgani}]{2009MNRAS.397..411T}
{Tescari} E., {Viel} M., {Tornatore} L., {Borgani} S., 2009, \mnras, 397, 411

\bibitem[{Theuns} et~al.(1998{\natexlab{a}}){Theuns}, {Leonard} \&
  {Efstathiou}]{1998MNRAS.297L..49T}
{Theuns} T., {Leonard} A., {Efstathiou} G., 1998{\natexlab{a}}, \mnras, 297,
  L49

\bibitem[{Theuns} et~al.(1998{\natexlab{b}}){Theuns}, {Leonard}, {Efstathiou},
  {Pearce} \& {Thomas}]{1998MNRAS.301..478T}
{Theuns} T., {Leonard} A., {Efstathiou} G., {Pearce} F.~R., {Thomas} P.~A.,
  1998{\natexlab{b}}, \mnras, 301, 478

\bibitem[{Tytler}(1987)]{1987ApJ...321...49T}
{Tytler} D., 1987, \apj, 321, 49

\bibitem[{Wiersma} et~al.(2009{\natexlab{a}}){Wiersma}, {Schaye}, {Theuns},
  {Dalla Vecchia} \& {Tornatore}]{2009MNRAS.399..574W}
{Wiersma} R.~P.~C., {Schaye} J., {Theuns} T., {Dalla Vecchia} C., {Tornatore}
  L., 2009{\natexlab{a}}, \mnras, 399, 574

\bibitem[{Wiersma} et~al.(2009{\natexlab{b}}){Wiersma}, {Schaye} \&
  {Smith}]{2009MNRAS.393...99W}
{Wiersma} R.~P.~C., {Schaye} J., {Smith} B.~D., 2009{\natexlab{b}}, \mnras,
  393, 99

\bibitem[{Wolfe} et~al.(2005){Wolfe}, {Gawiser} \&
  {Prochaska}]{2005ARA&A..43..861W}
{Wolfe} A.~M., {Gawiser} E., {Prochaska} J.~X., 2005, \araa, 43, 861

\bibitem[Zheng \& Miralda-Escud{\'e}(2002)]{2002ApJ...568L..71Z} 
Zheng, Z., \& Miralda-Escud{\'e}, J.\ 2002, \apjl, 568, L71 

\bibitem[Zwaan \& Prochaska(2006)]{2006ApJ...643..675Z} 
Zwaan, M.~A., \& Prochaska, J.~X.\ 2006, \apj, 643, 675 






\end{thebibliography}
\end{document}